# Fractal Fluctuations and Quantum-Like Chaos in the Brain by Analysis of Variability of Brain Waves: A New Method Based on a Fractal Variance Function and Random Matrix Theory

Elio Conte[(1,2)], Andrei Khrennikov[(3)], Antonio Federici[(1)], Joseph P. Zbilut [(4)]

[(1)]*Department of Pharmacology and Human Physiology and Tires, Center for Innovative Technologies for Signal Detection and Processing, University of Bari, Italy*
[(2)]*School of Advanced International Studies on Theoretical and nonLinear Methodologies-Bari, Italy*
[(3)] *International Center for Mathematical Modelling in Physics and Cognitive Sciences, M.S.I., University of Växjö, S-35195, Sweden*
[(4)]*Department of Molecular Biophysics and Physiology, Rush University Medical Center, 1653W Congress, Chicago, IL60612, USA.*

**Abstract:**
We develop a new method for analysis of fundamental brain waves as recorded by the EEG. To this purpose we introduce a Fractal Variance Function that is based on the calculation of the variogram. The method is completed by using Random Matrix Theory. Some examples are given.

## 1. Introduction

The electroencephalogram (EEG) is the electrical activity of an alternating kind generated from brain structures and recorded from the scalp surface after it is picked by metal electrodes and conductive media. If brain cells (neurons) are activated, local current flows are produced and EEG measures mostly the currents that flow during synaptic excitations of the dendrites of many pyramidal neurons in the cerebral cortex. Differences of electrical potentials are determined by summed postsynaptic graded potentials from pyramid cells. These create electric dipoles between soma and apical dendrites and brain electrical currents are due mostly to ion fluxes as $Na^+$, $K^+$, $Ca^{++}$, $Cl^-$ migrating through channels in neuron membranes as governed from membrane potential. At the microscopic level such processes, mainly unknown, must be seen in more detail taking in consideration the different types of synapses that involve a large variety of neurotransmitters.

The discovery of electrical currents in brain dates back to 1875 by R. Caton [1]. In 1924 H. Berger [1] measured brain's electrical activity and he was the first to describe brain electric potentials in humans outlining that brain activity changes in a recognizable manner when the general status of the subject changes as from relaxation to alertness.

In 1934 Adrian and Matthews were the first to discover the existence of "human brain waves [1], and in particular they identified regular oscillations around 10-12 Hz that they called "alfa rhythm".

It was the origin of research, studies and clinical applications about the presence of brain waves in humans.

We are accustomed today to analyze basic brain patterns of subjects by standard methodologies. Specifically, subjects are instructed to close their eyes and relax. Brain patterns are recorded as wave shapes that commonly sinusoidal. They are measured from peak to peak with a normal ranging from 0.5 to 100 $\mu V$. EEG records may be obtained by positioning 21 or more electrodes on the intact scalp and thus recording the changes of the electrical field within the brain. Generally, even up to 128 and more EEG channels can be displayed simultaneously and each corresponding to a standard electrode position on the scalp. The results of EEG signals are usually registered as voltage differences between pairs of electrodes with bipolar leads or between an active electrode and a suitably constructed reference electrode.

The problem in analyzing EEGs is to provide a proper method to extract its basic quantitative features by accurate procedures. The research regarding methodology began more than 70 years ago. The basic tool was, and still remains Fourier analysis. The brain states of subjects demonstrate some dominant frequencies; namely:
1) *beta waves (12-30Hz)*
2) *alpha waves (8-12 Hz)*
3) *theta waves (4-8 Hz)*
4) *delta waves (0.5-4 Hz)*

The argument of the present paper is of methodological nature. Over the last two decades the traditional Fourier analysis has been challenged by other methods, including the widespread application of time-frequency methods for signal analysis such as the Wavelet Transform (WT), and the Hilbert transform [2]. These applications have enjoyed varying results. Because of its simplicity, Fourier analysis has dominated and still dominates data analysis efforts. Despite this, it is widely recognized that the Fourier transform assumes crucial restrictions which are often violated in the EEG time series.[3]
In Fourier spectral analysis:
1) the system must be linear,
2) the data must be strictly periodic and stationary
3) all the data must be sampled at equally spaced time intervals; otherwise, at least the Lomb-Scargle elaboration should be used.

The consequences of improper FFT use are significant: the resulting spectrum will make little physical and physiological sense. Generally speaking, a spectrum may spread over a wide range of frequencies and we could estimate unawares spurious harmonics. An additional limit derives because Fourier spectral analysis holds only on linear superpositions of trigonometric functions. One introduces additional harmonic components any time a deformed wave profile is considered. Such deformations are the well known consequence of non linear contributions. In brief, we conclude that, following such standard methodology, the presence of non stationarity and of non linearity induce spurious harmonic components that give little sense to the results that are consequently obtained.

The brain has an average density of about $10^4$ neurons per cubic mm. Neurons are mutually connected into neural nets through synapses. Subjects have about 500 trillion ($5 \times 10^{14}$) synapses, and the number of synapses per one neuron increases with age while the number of neurons decreases with age. Thus although rather structurally simple, the interconnections produce one of the most massive (functional) structures existing in nature. The natural way to think of this structure is one of a dynamic system governed by laws of non linearity and of non stationarity. Consequently, any method of analysis must take these features into consideration. The present paper is devoted to the exposition of a such a new method for analysis of brain waves.

## 2. The Theoretical Basis of the Method
The concept of variability for a given time series should be clear. However, we give here a direct example of calculation in order to elucidate further the conceptual basis of the method.
Let us consider a series that, for the brevity of our exposition, w will consider constituted only by six consecutive values in equally spaced time intervals. We express the data in the following manner:

$$R_1, R_2, R_3, R_4, R_5, R_6. \tag{1}$$

The first time we select the time lag $h = 1$ and, using (1), we will have

$$(R_1 - R_2)^2 + (R_2 - R_3)^2 + (R_3 - R_4)^2 + (R_4 - R_5)^2 + (R_5 - R_6)^2 \tag{2}$$

that we indicate by $\gamma_1(h)$ and it will express the variability of the given time series at the resolution of time scale $h = 1$.

In order to calculate the final value of variability for such a time scale (or time-delay) we may decide to divide (2) by the number of pairs employed in this calculation and we will obtain the mean value of the variability at such time scale. Otherwise we could also divide by 2N, N being the total number of points in the given series, and in this case we will speak of total variability at the same prefixed resolution of time scale.

Note some important features:

1) The differences $(R_i - R_j)^2$ in (2) will account directly of the fluctuations that intervene in $R_{i+1}$ with respect to $R_i$

2) Still, the count of such variability will happen for all the points of the given time series and thus it will account for the whole considered dynamic process.

3) Finally, if $\gamma_1(h)$ will assume a value going to zero, we will conclude that at such resolution of the time scale (time lag delay $h = 1$) the process exhibits a quasi-periodic behaviour. Of course, we will consider that it gives great variability if $\gamma_1(h)$ is very different from zero.

We may now select the subsequent time scale resolution. This is to say, we have now to select the time lag $h = 2$. We will have that

$$(R_1 - R_3)^2 + (R_2 - R_4)^2 + (R_3 - R_5)^2 + (R_4 - R_6)^2 \tag{3}$$

which we indicate by, $\gamma_2(h)$ since it expresses the variability at the resolution time scale ($h = 2$). Again we calculate the mean value of variability or the total variability and still again we estimate fluctuations in the given time series and thus along the whole dynamic process. The basic difference is that this time we calculate at a different time scale of resolution(time lag delayed ) $h = 2$. Again a value of $\gamma_2(h)$ approaching zero will indicate a quasi–periodic behaviour while a very different value of $\gamma_2(h)$ from zero will indicate the presence of relevant variability.

In conclusion, we may calculate variability and quasi-periodic behaviours step by step at different time scales. The final sum of such calculated variability will give the value of the final variability exhibited by the process. The result will be a diagram in which the axis of the ordinate (usual y-axis) we express the values of variability, while instead in the axis of abscissa (x-axis) we give the corresponding values of *h*.

There is still a statement that must be added. For $\gamma_i(h)$ values approaching zero we have spoken of a quasi periodic regime. Our argument is strongly linked to the RQA methodology . More correctly, in Recurrence Quantification Analysis (RQA) [4] one speaks about Recurrences. Here we speak about variability. These two terms demonstrate very slight differences. The conceptual point is that the variability in a given time series has a "mirror" counterpart in terms of its recurrence. Quoting directly from Chapter (2), [4], one says that the common ground between living and non living systems resides in their shared property of recurrence. That is, within the dynamical signals expressed by living and non living signals are stretches, of repeating patterns and patterns of recurrence in nature have.

Variability vis-à-vis recurrence may be considered in the following way: the RQA method fixes the RQI method to count the percent number of recurrences given at various lags and within a predefined radius of convergence. Calculating RQI one obtains exactly the behaviour of a variogram overturned which we will introduce in the following equations. We will show that the variogram produces an accurate quantification of the $\gamma_i(h)$. In conclusion, our method demonstrates a link with RQI in Recurrence Quantification Analysis (RQA), but it offers also the possibility of making accurate numerical evaluations of the dynamics in the process that we have under investigation.

Finally, it should be also pointed out that we may also employ a reconstruction of the time series in phase space through embedding. In this case we account for variability of the signal in its proper phase space reconstruction. Therefore, in the present paper we are exposing the most simple version of the method but we may also utilize a more sophisticated analysis of variability of the given signal.

**The variogram technique**.

Various methods have been introduced to study the scaling properties of a given time series $X(t)$. Historically, one of the most effective is represented by the rescaled range statistical analysis that was first introduced by H.E. Hurst.

In essence, the Hurst analysis [5] reveals some statistical properties of a time series $X(t)$ scale with an observed period of observation $T$ and a time resolution $\mu$. Scaling results are characterized by a well known exponent $H$ that relates the long-term statistical dependence of the signal. In substance, the Hurst approach, expresses the scaling behaviour of statistically significant properties of the signal. In other terms, indicating by $E$ mean values, the problem becomes one analyzing the q-order moments of the distribution of the increments

$$K_q(\tau) = \frac{E(|X(t+\tau) - X(t)|^q)}{E(|X(t)|^q)} \tag{4}$$

Eq. (4) represents the statistical time evolution of a given stochastic variable $X(t)$.
For q = 2, we may re write (4) in the following manner [6]:

$$\gamma(h) = \frac{1}{2n(h)} \sum_{i=1}^{n(h)} [X(u_{i+h}) - X(u_i)]^2 \tag{5}$$

Eq. (5) estimates the variogram that we consider in the present paper. Here $n(h)$ is the number of pairs at lag distance $h$ while $X(u_i)$ and $X(u_{i+h})$ are time sampled series values at times $t$ and $t+h$), $t = u_1, u_2, \ldots$, $h = 1,2,3,\ldots$.

The variogram is thus a statistical measure in the form

$$\gamma(h) = \frac{1}{2} Var[X(u+h) - X(u)] \tag{6}.$$

We may re write this argument giving some more detailed mathematical characterization.
Let us consider X(t) to represent a time series (($t \in D$ that is a subset of $R_+$). As general scheme, let us suppose that this signal $X(t)$ is composed by the sum of a deterministic part, $\mu(t)$, plus a stochastic part. According to (4), (5), (6), we consider that the series satisfies the hypothesis given in (6) that we may re write in the following manner

$$Var.[X(t+h) - X(t) = 2\gamma(h)] \qquad \forall\ t,\ t+h \in D. \tag{7}$$

Note that through (5) we may perform also a fractal analysis of the given time series. To this purpose we use the Fractal Variance Function, $\gamma(h)$, the Generalized Fractal Dimension, $D_{dim}$, by the following equation

$$\gamma(h) = Ch^{D_{dim}} \tag{8}$$

and the simple marginal Density Function for self-affine distributions, given in the following manner

$$P(h) = ak^{-a}h^{a-1} \qquad (9)$$

with $D_{dim} = a - 1$ and $k$ the scale parameter.

An example of the methodology is given in [6].
According to [7] we may perform also multifractal analysis. A generalized Hurst exponent $H(q)$ may be defined starting with the scaling behaviour of $K_q(\tau)$ that may be rewritten as follows

$$K_q(\tau) \cong \left(\frac{\tau}{\nu}\right)^{qH(q)} \qquad (10)$$

with time resolution $\nu$.
If our analysis gives $H(q) =$ constant, that is to say, it is independent of $q$, we are consequently exploring a uni-scaling or uni-fractal process. In this case we have a unique constant $H$ that is the Hurst exponent. Instead, when $H(q)$ is not constant, and it depends on $q$, the process will be multi-scaling or, as it is often said, multi-fractal.
This last elaboration completes the exposition of our method.
*We reach here the central aim of the present paper based on the formulas from* (8)-(10) *by* (5).
The observed power law form, given in (8) and $1/f^{\beta}$ behaviour as expressed in (14), are assessable case by case starting with every EEG time series and calculating the variogram by (5). The possibility of observation and analysis of a power law and $(1/f^{\beta})$ behaviour in the EEG is of fundamental interest. Case by case it indicates that we may be viewing fractal fluctuations in the EEG. Such fractal fluctuations may be indicative of long range space-time correlations. Still, the presence of a universal power law could be also indicative of self-organization or, specifically, of self- organized criticality. The presence of a power law as well as of a fractal dimension measure may be seen in relation with the presence of quantum or quantum like dynamics and chaos in brain. To this purpose we point out that in our previous studies we reached evidence of the presence of quantum like dynamics in brain and in cognitive entities [8]. By the present methodology we are now in the condition to extend the quantum possible analysis of such brain dynamics. Therefore, this is the major feature of our methodology here discussed.
We have now reached the goal of the proposed method; that is, to study the variability of EEGs, and, generally speaking, of a given time series that exhibits periodic or pseudo-periodic behaviours looking also to possible quantum or quantum like presence in such systems. Obviously, the basic feature of the CZF method is that by it we can estimate the variability that one has in the EEG for each band in a given time interval and this represents the new and important feature of the method. By CZF the new key in EEG analysis becomes one of variability in microvolts in a given time. In section 3 which is dedicated to some applications of CZF we will present some results. For one subject, just to give a clear indication on the state of art, we will include also a Recurrence Quantification analysis that was performed by us on the 60000 EEG points analyzed by epochs. We selected 600 epochs and thus having each epoch of 100 points corresponding to 0.4 sec. We report such results for %Recurrences, % Determinism, %Laminarity, Entropy, Max Line, Trapping Time. As it is well known, the variable %Rec. must be intended in relation to the quasi periodicity that our EEG signal still exhibits. %Det. instead characterizes still the determinism of the investigated signal. %Lam represents fast transitions (for example, chaos-chaos transitions) or changes from one state to another or intermittency and instabilities. Trapping Time indicates instead the actual time spent in the transition. Still, Entropy may be seen in relation to the complexity of the investigated signal, while the inverse of MaxLine results to be proportional to the local estimation of the dominant Lyapunov exponent.

It is useful to observe such dynamical behaviours in the case of our investigated EEG to convince ourselves about the high level of diversity that the brain dynamics exhibits and consequently to show that the basic concept we must use in analysis of brain waves of EEG is that one of variability through CZF. This seems the most correct way to investigate brain dynamics, fractal fluctuations, long range correlations and the possible signature of quantum or quantum-like chaos.

In order to complete our exposition we must now give indication on the manner in which an analysis of brain waves must be performed using the variogram starting from the recorded EEG. Let us give an example to be clear.

Consider an EEG sampled at 250 Hz . First of all we will calculate the variogram for different lags, $h$, as indicated in (5) and previously explained in detail. We will realize a diagram in which we have the values of the variogram in y-axis (ordinate) and correspondingly the $h-lag-values$ on the x-axis (abscissa). Our problem is now a conversion of variogram values to values of frequency ($Hz$). We proceed as it follows.

$$\frac{1}{250\ Hz} = 0.004\ \sec. \tag{11}$$

In this manner

$$\frac{1}{0.004 \times lag - h - value = 1} = 250\ Hz \tag{12}$$

will represent the frequency with the corresponding value of variability at $250\ Hz$. Similarly,

$$\frac{1}{0.004 \times 2\ (lag\ value)} = 125\ Hz \tag{13}$$

will represent the value of variability at $125\ Hz$, and so on for lag values $h = 3,4,5,......$. In this manner we may reconstruct the variability of the EEG time series data as a function of the frequency.

Analysis of brain waves will be performed by integration of the calculated variability in each of the four groups of brain waves previously reported summing for each characteristic frequency band.

With respect to the (8) and (9), on the basis of the (11-13) an

$$P(f) = 1/f^{\beta} \tag{14}$$

behaviour can be also calculated and estimated.

For the search of significant correlations we will use Random Matrix theory [9].

Let us indicate by $x(t)$ ($t = 1,2,......,T$) the given time series of the data. Having fixed the dimension $m$ of the reconstructed space for such observations, we will write the matrix $Z$ of the delayed observations such that the first column of $Z$ will be $x(t)$, the second $x(t-1)$, and so on to $x(t-m)$ for the $(m+1)th$ column. The correlation matrix $C$ will be given by the following equation

$$C = \frac{Z^T Z}{T-m} \tag{15}$$

At this stage we will ascertain deviations of properties of $C$ from the correlation matrix of random series. Let us remember that for a random matrix $M \times N$, the range of the eigenvalues of its correlation matrix is include between a minimum and maximum eigenvalues, given respectively in the following manner

$$\lambda_{minimum} = \sigma^2 \left(1 - \frac{\sqrt{k}}{k}\right)^2 \tag{16}$$

and

$$\lambda_{maximum} = \sigma^2 \left(1 + \frac{\sqrt{k}}{k}\right)^2 \tag{17}$$

where $\sigma^2$ represents the variance of the elements in the correlation matrix and $k = \dfrac{M}{N}$. The (16) and the (17) are valid in the limit of a very large value of $k$. The remaining step is to ascertain the presence of a null hypothesis for a random matrix. The eigenvalues of $C$ must be compared with the range of eigenvalues of the correlation matrix of a purely random matrix as fixed by the (16) and the (17). If the computed eigenvalues result within the theoretical limits of those of a random matrix we will conclude for any correlations with no real significance.

This last discussion completes the exposition of our method. It will be applied to EEG as well as to ERP.

We will call it the CZKF, recalling the first letters in the surnames of the authors.

**3. An Example of Application**

We examined eight normal subjects (5 female and 3 male with age ranging from 21 to 28 years old) All the subjects were at rest, watchful but with closed eyes .The sampling frequency was at 250 Hz. We focused our analysis on the following electrodes: $C_Z$, $F_Z$, $O_2$, and $T_4$.

Phase space reconstruction is useless in our case since we had the electrodes positioned on the scalp and their space separation corresponds to time delay. We used Euclidean Norm that is the time series reconstructed as

$$\sqrt{x_{C_Z}^2(t) + x_{F_Z}^2(t) + x_{O_2}^2(t) + x_{T_4}^2(t)} = X_{EEG}(t) \quad (18)$$

and we calculated the variogram of $X_{EEG}(t)$ at the various lags as previously explained, and converted into $Hz$ as in (11), (12), and (13). In (18) 30000 points of EEG were used, corresponding to 2 minutes of recorded brain activity. Every subject was analyzed two times, (cases A and B in the table) with a difference of several days between analyses.

The results are reported in Fig.1 and Table 1 were it is seen that the CZF method enables us to analyze in detail the behaviour of brain waves in EEG with accuracy, accounting in particular for all the contributions arising during the brain dynamics and including in particular the contributions of non linear and non stationary components . The variogram given in Fig.1 is estimated every 0.004 sec. that is roughly the refractory time of the neuron. Its values may be subsequently used to perform fractal analysis by (8), (9), (10), (14) and to evaluate fractal fluctuations, and thus possibly self-organized criticality and quantum-like chaos in the macroscale dynamical system investigated. In this paper we performed calculations of generalized fractal dimension as given in (8) and the results are given in Table2. We also provide a plot of the results for some subjects. The other graphs are very similar. Note that in Fig.2 we have the results of RQA analysis for a subject but the results did not change substantially, as neither did the other remaining subjects. By this analysis we evaluate the overall complexity of our investigated system. We estimate its Recurrence that is the quasi-periodicity that the systems still maintains. We also estimate its level of Determinism, its complexity by entropy, its local Lyapunov exponents by measuring the inverse of Max Line, and finally we characterize the Laminarity that expresses the transitions chaos-chaos and/ chaos/order that we have in the system, and, finally, the Trapping Time that is an estimation of the time the system remains in its state before executing a new transition. This time the analysis was performed on 60000 points - EEG data with 600 epochs, each epoch of 100 points , that is to say about 0.4 sec. A complete fractal analysis pertaining in particular multifractal investigation, could  be now performed on the basis of the (8), (9) and (10) by the CZF method. This is the aim of the subsequent part II of the present paper.

Table 1 CZF: Analysis of brain waves

| Subject name | delta <4 Hz | 4<teta<8 Hz | 8<alfa<12 Hz | 12<beta<30 Hz | 30<gamma<50 Hz | 50-125 Hz |
|---|---|---|---|---|---|---|
| AM. A | 315830.18 | 1546.41 | 512.81 | 511.46 | 124.08 | 40.96 |
| AM. B | 345604.54 | 1537.95 | 485.86 | 564.77 | 158.55 | 65.03 |
| CR. A | 342601.77 | 1533.87 | 593.84 | 992.27 | 236.30 | 100.41 |
| CR. B | 231064.75 | 1184.40 | 360.35 | 439.72 | 135.65 | 50.88 |
| CB. A | 269108.24 | 1412.73 | 477.23 | 497.33 | 143.51 | 69.65 |
| CB. B | 438748.26 | 2268.66 | 775.45 | 781.64 | 206.83 | 96.67 |
| DPC. A | 1157817.77 | 5349.84 | 1487.23 | 1438.89 | 395.36 | 181.99 |
| DPC. B | 770427.70 | 3858.46 | 1096.07 | 1095.69 | 335.10 | 127.57 |
| IC. A | 296854.43 | 1635.37 | 561.97 | 592.92 | 177.03 | 107.97 |
| IC. B | 420348.11 | 2272.28 | 769.46 | 799.45 | 243.02 | 135.93 |
| RR. A | 462992.76 | 2266.45 | 694.40 | 773.47 | 264.10 | 105.22 |
| RR. B | 855793.00 | 3727.05 | 1128.16 | 1258.53 | 474.90 | 209.62 |
| SC. A | 625474.63 | 2916.07 | 871.99 | 882.63 | 234.88 | 108.78 |
| SC. B | 430362.95 | 2232.10 | 735.97 | 829.91 | 261.08 | 123.09 |
| VD. A | 979082.92 | 4177.67 | 1128.36 | 1435.87 | 481.32 | 175.21 |
| VD. B | 882707.17 | 3041.39 | 986.53 | 1046.34 | 355.57 | 146.65 |

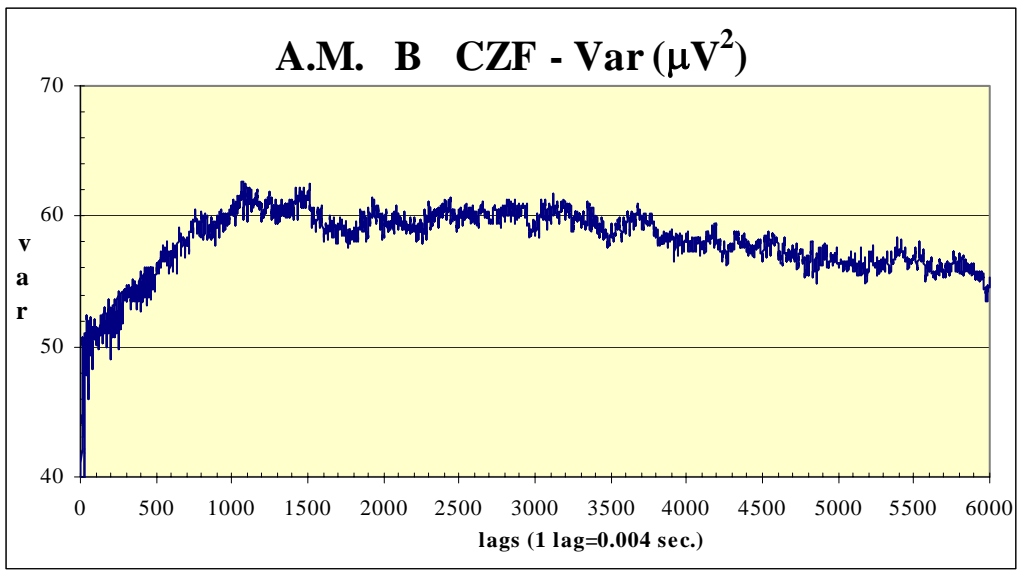

**Fig. 1**

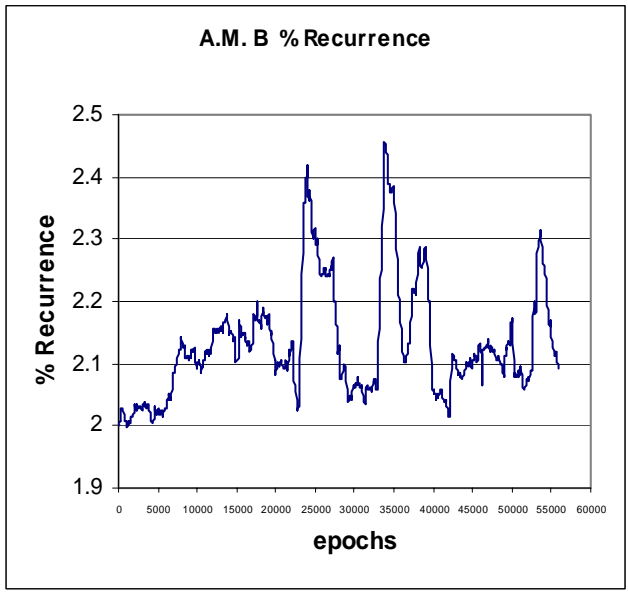
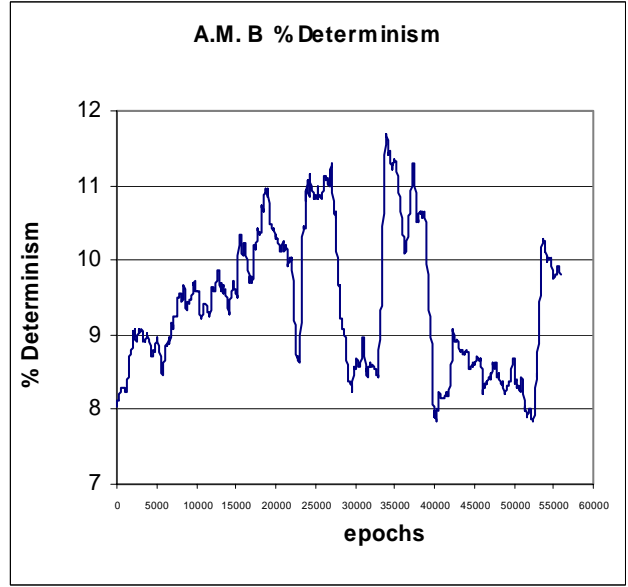
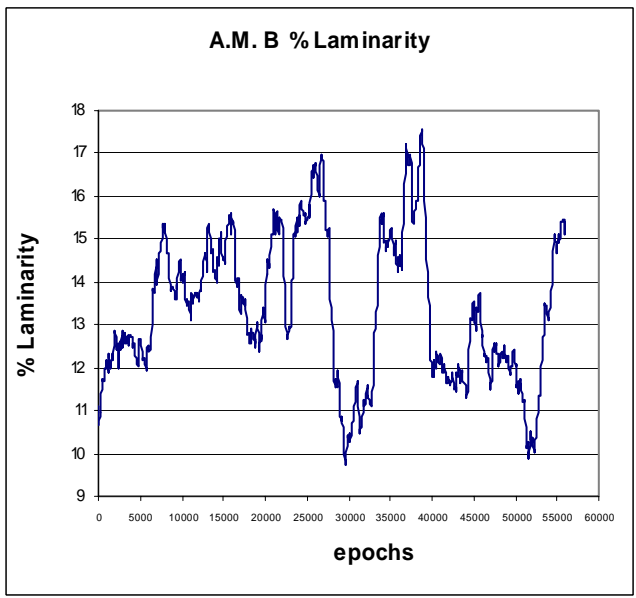
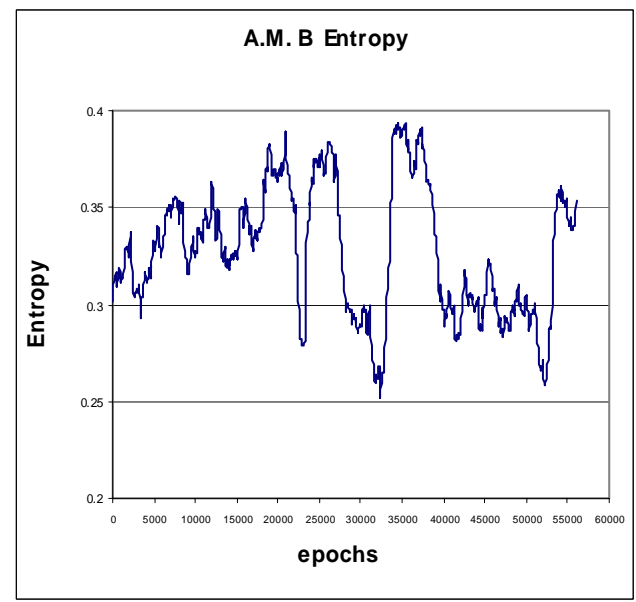
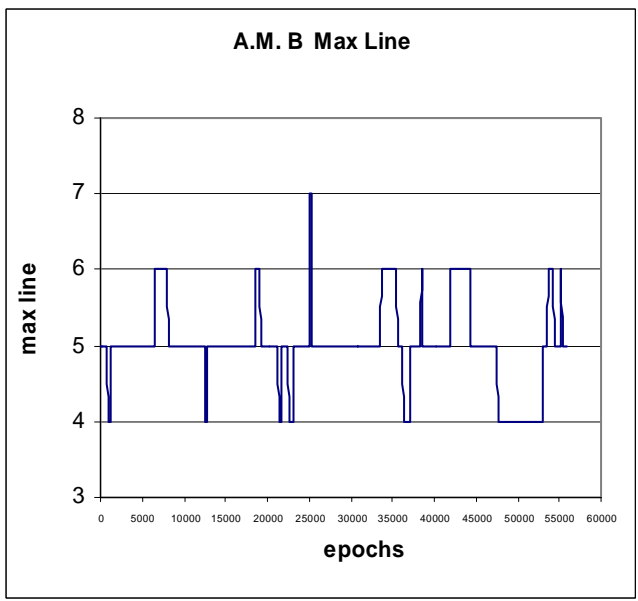
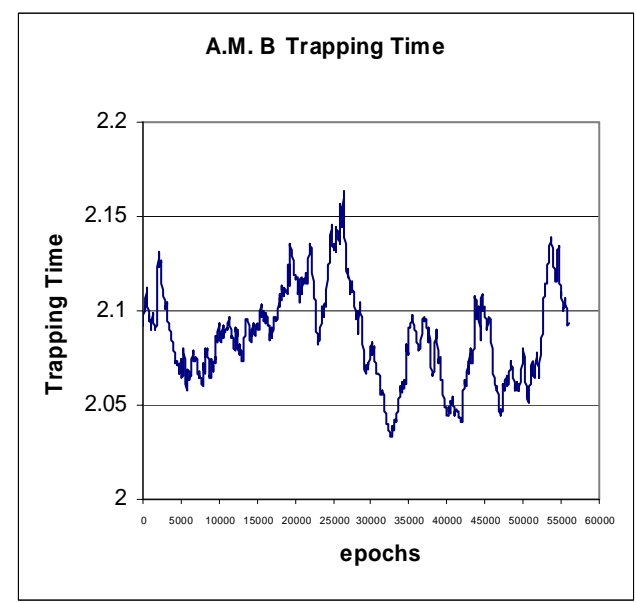

**Fig. 2**

## Tab. 2: Calculation of Generalized Fractal Dimension : $\gamma(h) = C\, h^D$

### Model Estimation Section AM: -A

| Parameter Name | Parameter Estimate | Asymptotic Standard Error | Lower 95% C.L. | Upper 95% C.L. |
|---|---|---|---|---|
| C | 23.53934 | 1.216799 | 21.12465 | 25.95404 |
| D | 0.1858372 | 1.306907E-02 | 0.1599021 | 0.2117724 |

### Model Estimation Section: AM - B

| Parameter Name | Parameter Estimate | Asymptotic Standard Error | Lower 95% C.L. | Upper 95% C.L. |
|---|---|---|---|---|
| C | 29.07217 | 1.396714 | 26.30044 | 31.84391 |
| D | 0.13293 | 1.229815E-02 | 0.1085247 | 0.1573353 |

### Model Estimation Section: CR -A

| Parameter Name | Parameter Estimate | Asymptotic Standard Error | Lower 95% C.L. | Upper 95% C.L. |
|---|---|---|---|---|
| C | 9.162621 | 0.4410343 | 8.287403 | 10.03784 |
| D | 0.1841901 | 1.130636E-02 | 0.157753 | 0.2106272 |

### Model Estimation Section: CR -B

| Parameter Name | Parameter Estimate | Asymptotic Standard Error | Lower 95% C.L. | Upper 95% C.L. |
|---|---|---|---|---|
| C | 22.3534 | 1.119441 | 20.13191 | 24.5749 |
| D | 0.1362431 | 1.280879E-02 | 0.1108245 | 0.1616617 |

### Model Estimation Section: DPC - A

| Parameter Name | Parameter Estimate | Asymptotic Standard Error | Lower 95% C.L. | Upper 95% C.L. |
|---|---|---|---|---|
| C | 51.82169 | 1.685851 | 48.47618 | 55.16721 |
| D | 0.3080554 | 8.035804E-03 | 0.2921086 | 0.3240022 |

### Model Estimation Section : DPC -B

| Parameter Name | Parameter Estimate | Asymptotic Standard Error | Lower 95% C.L. | Upper 95% C.L. |
|---|---|---|---|---|
| C | 46.12198 | 2.012183 | 42.12887 | 50.11509 |
| D | 0.2479528 | 1.089153E-02 | 0.2263389 | 0.2695667 |

### Model Estimation Section: IC - A

| Parameter Name | Parameter Estimate | Asymptotic Standard Error | Lower 95% C.L. | Upper 95% C.L. |
|---|---|---|---|---|
| C | 37.71668 | 0.8622698 | 36.00553 | 39.42782 |
| D | 8.428948E-02 | 5.928888E-03 | 0.0725238 | 9.605516E-0 |

## Model Estimation Section: IC -B

| Parameter Name | Parameter Estimate | Asymptotic Standard Error | Lower 95% C.L. | Upper 95% C.L. |
|---|---|---|---|---|
| C | 44.58364 | 1.407745 | 41.79002 | 47.37727 |
| D | 0.1250779 | 8.098834E-03 | 0.109006 | 0.1411498 |

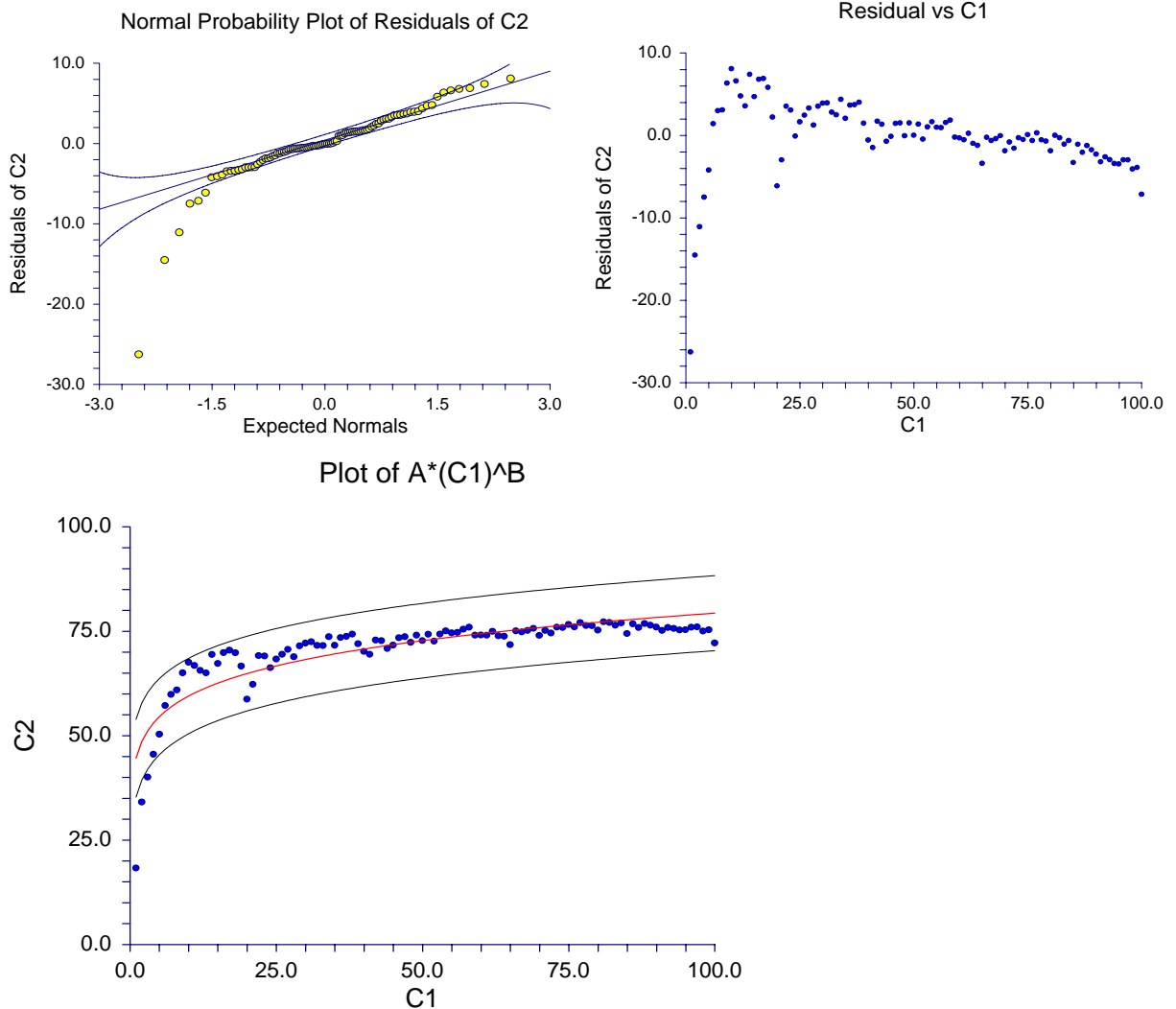

## Model Estimation Section RR: - A

| Parameter Name | Parameter Estimate | Asymptotic Standard Error | Lower 95% C.L. | Upper 95% C.L. |
|---|---|---|---|---|
| C | 38.77415 | 1.74751 | 35.30627 | 42.24202 |
| D | 0.1640375 | 1.145067E-02 | 0.141314 | 0.186761 |

## Model Estimation Section: RR -B

| Parameter Name | Parameter Estimate | Asymptotic Standard Error | Lower 95% C.L. | Upper 95% C.L. |
|---|---|---|---|---|
| C | 63.69815 | 3.798292 | 56.16056 | 71.23573 |
| D | 0.1689539 | 1.513296E-02 | 0.138923 | 0.1989847 |

## Model Estimation Section: SC - A

| Parameter Name | Parameter Estimate | Asymptotic Standard Error | Lower 95% C.L. | Upper 95% C.L. |
|---|---|---|---|---|
| C | 34.47267 | 1.386476 | 31.72125 | 37.22409 |
| **D** | **0.2519439** | **1.003323E-02** | **0.2320333** | **0.2718545** |

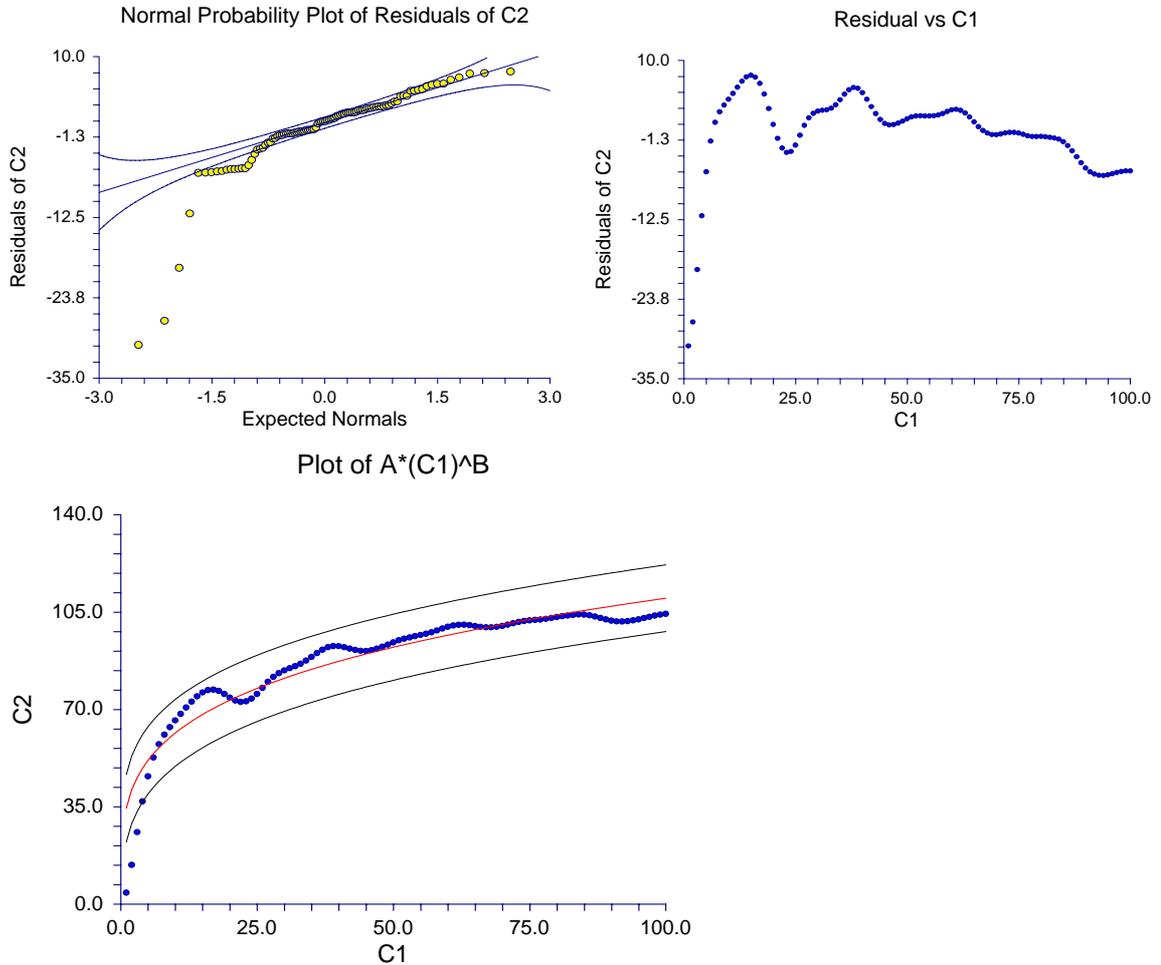

## Model Estimation Section: SC - B

| Parameter Name | Parameter Estimate | Asymptotic Standard Error | Lower 95% C.L. | Upper 95% C.L. |
|---|---|---|---|---|
| C | 42.99538 | 2.115731 | 38.79678 | 47.19398 |
| **D** | **0.1299787** | **1.260582E-02** | **0.1049629** | **0.1549945** |

## Model Estimation Section: VD - A

| Parameter Name | Parameter Estimate | Asymptotic Standard Error | Lower 95% C.L. | Upper 95% C.L. |
|---|---|---|---|---|
| C | 66.81789 | 4.375697 | 58.13446 | 75.50132 |
| **D** | **0.1781886** | **1.658479E-02** | **0.1452767** | **0.2111006** |

### Model Estimation Section: VD - B

| Parameter Name | Parameter Estimate | Asymptotic Standard Error | Lower 95% C.L. | Upper 95% C.L. |
|---|---|---|---|---|
| C | 50.59438 | 2.474958 | 45.68291 | 55.50586 |
| **D** | **0.1777683** | 1.238973E-02 | 0.1531812 | 0.2023553 |